# Reconfigurable single photon sources based on functional materials

*Dmytro Kundys\*, Francesco Graffitti, Richard A. McCracken, Alessandro Fedrizzi and Bohdan Kundys*

Dr. Dmytro Kundys, Francesco Graffitti, Dr. Richard A. McCracken, Dr. Alessandro Fedrizzi
Scottish Universities Physics Alliance (SUPA),
Institute of Photonics and Quantum Sciences, Heriot-Watt University
Edinburgh EH14 4AS, UK
E-mail: d.kundys@hw.ac.uk

Dr. Bohdan Kundys
Institut de Physique et Chimie des Materiaux de Strasbourg (IPCMS),
UMR 7504 CNRS-UdS 23 Rue du Loess,
67034 Strasbourg, France

The future of quantum photonic technology depends on the realization of efficient sources of single photons, the ideal carriers of quantum information. Parametric downconversion (PDC) is a promising route to create highly coherent, spectrally pure single photons for quantum photonics using versatile group-velocity matching (GVM) and tailored nonlinearities. However, the functionality to actively control the poling period of nonlinear crystals used in PDC is currently missing, yet would enable to dynamically modify the wavelength of single photons produced in the PDC process. Here a detailed GVM study is presented for functional PMN-0.38PT material which can be dynamically repolled at ambient conditions with fields as low as 0.4 kV/mm. Our study reveals phase-matching conditions for spectrally pure single photon creation at 5 – 6 microns. Further, a practical approach is proposed for on-flight wavelength switching of the created single photons. The reported reconfigurable functionality benefits a wide range of emerging quantum-enhanced applications in the mid-IR spectral region where the choice of single photon sources is currently limited.

## 1. Introduction

In recent years quantum technology has shown immense progress in demonstrating a variety of photonic platforms allowing coherent quantum control using pure single photon PDC sources[1,2], quantum dots[3,4] and diamond NV centers[5,6]. Harnessing single photons for quantum information encoding stands out as a promising direction, which on one hand benefits from superior coherence robustness of the photons at ambient temperatures, and on the other from the maturity of field of photonics in general, allowing numerous ways for the single photons to be generated, manipulated and detected. Single photon sources, therefore, constitute a crucial part of today's quantum photonics agenda. It has been shown that despite probabilistic operation, PDC sources can produce highly pure single photons approaching on-demand operation in multiplexed schemes[7]. There has been a surge of interest in single photon sources from a nonlinear parametric processes operated in wider spectral regions[8] and in mid-IR in particular[9–11]. The applications for mid-IR single photon sources primarily concern quantum sensing and metrology[12–14], stealth ranging and quantum LIDARs[15–17], quantum enhanced medical imaging[18–20], as well as for free-space secure communication in the atmospheric window, in light of recent demonstration of entanglement distribution using satellite-to-ground downlinks[21].

In the nonlinear PDC process, one pump photon is decomposed into the two photons (signal and idler) satisfying energy and momentum conservation. While there is little problem with preserving the energy conservation, it proves to be harder to satisfy the momentum conservation due to the natural dispersion of the nonlinear crystal. Techniques such as quasi-phase matching via periodic poling of ferroelectric nonlinear crystals allow to compensate the wave vector mismatch by adding additional wave vector $k_g = 2\pi/\Lambda$ of the poling grating, where $\Lambda$ is a poling period. All current periodically poled ferro-electric crystals used for PDC photon sources, primarily $LiNbO_3$[22] and $KTiOPO_4$[23–25], lack tunability of the poling period once fabricated, making the spectral characteristics of the created photons inaccessible for an active reconfigurability. Tailoring the poling period enables the option to change the wavelengths with a single crystal, switch the PDC generation on and off, or to reshape wavepackets of the subsequent photons. In conventional materials it requires large coercive fields and high temperatures to modify the crystal domain structure (see **Table 1**). In this letter we introduce a material that on one hand can be dynamically polled, and on the other suitable for the generation of spectrally pure single photons. We find this unique combination in the ferroelectric lead magnesium niobate-lead titanate, $(1-x)Pb(Mg_{1/3}Nb_{2/3})O_3 - xPbTiO_3)$ or short PMN-PT and show it is possible to obtain *both* pure single photon *and* wavelength-switching operation in the first atmospheric window, where there is currently a scarcity of single photon sources[9].

Unlocking dynamic access to the poling period allows active shaping of the spectral and temporal properties of the single photons [1], for example to apply or remove apodization on demand, to optimize single photon purity, or to manipulate the single photon wave-packet at a particular wavelength. Reconfigurable switching functionality, in particular, is very promising for free-space quantum secure communication as it permits to change the frequency band of the communication channel, which has not been shown previously with the PDC sources.

We focus on the *x*=38% composition (PMN-0.38PT), which exhibits a wide transparency region[26], high refractive index, low coercive field, and low Curie temperature $T_c$ of just 180°C, enabling functional poling domain switching at ambient temperatures using <1 kV/mm fields[27], which is not possible with other conventional nonlinear media used for PDC. We provide a comparison of some key material properties in Table 1.

The unique ferroelectric switching mechanism in PMN-PT is realized thanks to the ultrarich phase diagram where different thermodynamically equivalent crystalline phases manifest metastable coexistence in close to morphologic phase boundary region[28]. This is ideal for external domain poling control, where 180° polarization switching can be realized by weak electric fields[27]. Despite being known as highly nonlinear material[29,30] with strong electro-optical coefficients[31–33] the reports on nonlinear optical properties remain limited to just second harmonic generation[34]. To the best of our knowledge, the periodic poling has not been demonstrated for nonlinear PDC in bulk PMN-PT materials. However, a microscale periodically poled PMN-PT with precise domain structure down to 5 microns has been demonstrated by electron beam patterning technique[30] and also 200 nm pitch size by backswitch poling using nanopatterned composite electrodes[35]. The main issues with PMN-PT periodic poling has been the formations of cracks but very recently, there has been an important progress made in achieving high optical quality monodomain state in PMN-PT by using pre-poling thermal annealing[29] to prevent cracking which is an important step for nonlinear optical applications of PMN-PT material family. We report the first study which describes, evaluates and proposes a practical scheme of dynamic poling of nonlinear crystal by using a single electrode mask and show a variety of PDC types which are possible to achieve for PMN-0.38PT.

## 2. Parametric down conversion and group velocity matching

The material composition of PMN-0.38PT is an ideal candidate for mid-IR GVM because it offers a wide optical transparency window extending up until 6 microns. Normally GVM[36] is employed to maximize heralded-photon purity by minimizing spectral correlations of the first-order PDC bi-photon state:

$$|\Psi_{pair}\rangle = \iint d\omega_s\, d\omega_i f(\omega_s, \omega_i)\hat{a}_s^\dagger(\omega_s)\hat{a}_i^\dagger(\omega_i)|0\rangle_{s,i} \qquad (1)$$

where $f(\omega_s,\omega_i)$ denotes the joint spectral amplitude (JSA) which is determined both by spectral bandwidth of the pump photon and the nonlinear profile of the crystal, and $\omega_{s,i}$ denotes the signal and idler photon frequencies respectively. In this letter we consider both degenerate and nondegenerate cases where signal and idler are of the same and different wavelengths respectively. The refractive index Sellmeier coefficients were taken from He et al.[36] using the Wemple–DiDomenico single oscillator dispersion model[37] for mid-IR wavelength range dispersion approximation, along with thermal expansion data from[38].

Pure single photon source engineering via GVM lies in selecting conditions under which the pump envelope function (PEF) $\alpha(\lambda_s, \lambda_i)$ and nonlinear crystal phase-matching function (PMF) $\Phi(\lambda_s, \lambda_i, T)$ can form close-to-separable JSA of the bi-photon state:

$$f(\omega_s, \omega_i, T) = \alpha(\omega_s, \omega_i) \times \Phi(\omega_s, \omega_i, T) \qquad (2)$$

On the other hand, an optimal GVM condition can be estimated when the dispersion parameter $D = -(GD_p - GD_s)/(GD_p - GD_i)$ is $D > 0$, where $GD$ represent group delays of the three interacting photons pump, signal, and idler respectively. The dispersion parameter allows the evaluation of the tilt of the phase-matching function, $\theta = tan^{-1}(D)$, in the signal-idler wavelength space, with respect to the *x*-axis [1,39].

In principle the best single photon purity can be obtained when the PMF tilt angle $\theta$ is anywhere between 0 and 90 degrees. High signal-idler indistinguishability can be achieved when $\theta = 45$, a condition known as symmetric group velocity matching, when PEF and PMF are aligned orthogonally thus forming a symmetric JSA profile. Moreover, photon purity can be optimized by apodizing the poling structure of the crystal to reduce the spectral correlations between the PDC photons[1].

To the best of our knowledge, PMN-0.38PT has not been studied previously for PDC, therefore we perform a detailed investigation of all possible collinear configurations of photon wavelengths and polarization. In our GVM calculations we use the following input parameters: pump center wavelength of 2.7 and 2.88 microns with FWHM of 35 nm, with $sech^2$ intensity profile corresponding to ~ 220 fs long transform limited pulses with bandwidth of just under 6 meV or 1.44 THz; the length of the crystal of 25 mm, chosen to match the spectral width of the PEF. We numerically solve the phase matching condition using the dispersion equation $n^2 = 1+S_0\lambda_0 /(1-\lambda_0/\lambda)$, where $S_0$, $\lambda_0$ are average oscillator strength and average interband oscillator wavelength in the long wavelength approximation with $S_{0o} = 1.004\times10^{14}$ m$^{-2}$, $\lambda_{0o} = 226\times10^{-9}$ m, $S_{0e} = 1.017\times10^{14}$ m$^{-2}$, $\lambda_{0e} = 223\times10^{-9}$ m for the ordinary and extra-ordinary orientations respectively[36].

In our code we also account for thermal expansion of the poling period with the coefficient of $3.8\times10^{-6}\,°C^{-1}$ extrapolated for the 38% PMN-PT from ref[38]. Based on the abovementioned parameters and the crystal temperature of 35°C we obtain the quasi-phase-matching under which the wavevector mismatch is made to vanish: $\Delta k = k(\lambda_p) - k(\lambda_s) - k(\lambda_i) - 2\pi/\Lambda = 0$, where $\Lambda$ is the crystal poling period.

We first consider the type-0 case with the two options $e \rightarrow e+e$, $o \rightarrow o+o$, further in text denoting pump conversion into signal and idler photons, where $e$ and $o$ indicate photon polarization along extra-ordinary and ordinary crystal axes respectively. We obtain phasematching results with $\Lambda$ of 0.493 mm and 0.510 mm for $o \rightarrow o+o$ and $e \rightarrow e+e$ cases respectively. The key observation from **Figure 1** is that the PMF is inclined at the same angle as PEF in Figure 1(*a-b*), which is not useful for pure photon generation[1]. From the numerical estimation of the value of $\theta$, the tilt of the phase-matching function, we also find no useful type-0 PDC condition for the wider span of the pump wavelengths ranging from 0.6 - 9 μm.

In the type I case, the two configurations $e \rightarrow o+o$ and $o \rightarrow e+e$, yield phasematching with $\Lambda = 0.297$ mm and 0.136 mm respectively. In **Figure 2** the GVM plots for the $e \rightarrow o+o$ configuration reveals a distinct feature with a large crossover in the PMF. Such spectrally broad PMF forms a JSA of a bi-photon spreading between 4.8 and 6 microns (Figure 2 (b)).

The behavior of the singularity observed in PMF can be further investigated at different pump wavelengths by looking into the GVM mismatch angle results presented in Figure 2 (c). The spectral GVM mismatch angle plots clearly show that singularity in $\theta$ occurs right at the degenerate point which in its turn hinders degenerate PDC condition needed for high purity photon generation. For the nondegenerate case, however, the GVM results still offer many interesting variations for $\theta$ between 0 and 90 degrees at the pump ranging between 1.5-3 μm (Figure 2 (c)) which can find its use in applications not requiring spectrally balanced single photon sources.

Finally, we obtain the most valuable result when investigating the type II PDC with the $e \rightarrow o+e$ configuration; and find that degenerate GVM is indeed possible for PMN-PT, with PMF inclination approximating 0 degrees for high purity bi-photon state in **Figure 3**. This result opens the possibility for PMN-PT class of materials to be accessed for high-purity single photon creation in the mid-IR spectral region. We further perform wider spectral range numerical simulation in order to find all possible phase-matching conditions. The results in Figure 3 (c) indicate strict $\theta = 0$ condition starts from 2.88 μm of the pump which correspond to signal/idler photons created at 5.76 μm which still fits into PMN-0.38PT transmission window far edge. It is important to note that that it is possible to obtain high purities at any PMF angle between 0 and 90 degrees by transforming the PMF from *Sinc* into a *Gaussian* profile by using domain engineering techniques[40]. The results in Figure 3 (c) show that when including the nongenerated cases it is possibility of generating pure bi-photon states at much larger spectral window spanning from 1 to 6 microns (limited by the crystal transmission edge).

Another degenerate PDC case is also available for the symmetric type II $e \rightarrow e+o$ case in **Figure 4**. While the critical spectral positions for the angles of $\theta = 0$, 90 and 45 for the both degenerate signal/idler cases are identical, when looking into the broad spectral range numerical modelling for the $\theta$ parameter we obtain slightly different outcome to the $e \rightarrow o+e$ for the non-degenerate PDC in Figure 3 and Figure 4, which allows extra flexibility for the crystal design parameters.

It is important to note that type II gives rather long poling periods with $\Lambda$ of 1.429 mm both for $e \rightarrow o+e$ and $e \rightarrow e+o$ cases. The implications of these long poling periods are two-fold: (i) firstly, it means that the design of the crystal length will have to take this into account when matching the bandwidth of the pump photons; (ii) secondly, and more importantly, long poling periods make it technologically easy to exploit the ferroelectric switching functionality by applying fractional modifications to the poling periods or perform sub-coherence domain engineering.

The switching functionality of the poling order can be realized by applying an electrode-patterned mask along the crystal length, thus allowing selective altering of the poling period of each particular domain on-flight. As an illustrative example of applying such functionality, we take a situation where we need to change the band of the communication channel in a mid-IR free-space quantum key distribution protocol by switching the wavelength of the created single photons from 5.4 μm to 5.6 μm. Provided we have two pump wavelengths of 2.7 and 2.8 μm available from a tunable OPO, such switching will require modifying the crystal poling period $\Lambda$ from 1.429 to 1.239 mm as per our calculations, i.e. reduced by 190 μm. With the functional PMN-PT material studied here, this modification is indeed possible owing to low coercive fields for on-flight domain reorientation as discussed earlier. We illustrate practical realization of this concept in **Figure 5**. Let us consider the initial crystal with the poling period of 1.429 mm (Figure 5 (a)). By applying a pre-fabricated electrode mask with a linear gradient elongation starting from the first electrode length and moving further along the waveguide, one can effectively shrink the poling period by then applying the poling voltage as shown in Figure 5 (c) therefore achieving the desired period. The effect of the domain boundary regions is negligible in our calculations due to their atomically thin volume fraction [41] in respect to the domain sizes which are of a millimeter order.

The switching process can be explained as follows. We start from the crystal in the Figure 5(a) top, with the electrode mask applied, and the poling period corresponding to 5.4 μm photon creation mode. The electrode mask is positioned such that when the reverse voltage is applied each electrode pair which is deposited at the edge of the corresponding domain will shrink each such domain by the length equal to the half of the period difference needed for the two wavelengths conversion 5.4 and 5.6 μm respectively, i.e. 95 μm. The length of each consecutive electrode pair has the length is incremented by N×95 μm where N is the number of the next domain, to ensure uniform poling period change across the entire crystal when the voltage is applied to reverse the poling orientation of the relative domain fraction. Moreover, the proposed design of the switching mechanism is fully reversible, and the previous poling period can be recovered simply by switching the polarity of the same electrode mask. The rate of dynamic domain repolling can potentially reach down to a *ns* scale and typically have endurance between $10^{10}$ and $10^{14}$ cycles[42]. Figure 5 (b) illustrates what effect the above-mentioned poling period modification has on the JSA's for each case. From the analysis of the JSA we can confirm obtaining a distinct and well separated JSA's of the bi-photon states. In principle, the proposed method can be applied further to achieve wider signal/idler wavelengths range by choosing suitable electrode mask for the corresponding poling period dependence shown in Figure 5 (c).

Moreover, single photon purity produced in PMN-PT can be further increased by *either* use of narrowband filtering if the tilt of the PMF is not ideal, *or* by the domain apodisation techniques[1] where the advantage of switching functionality of PMN-PT makes them ideal candidates for dynamic control of single photon spectral purity or even creation of frequency encoded photons for hyperentangled state generation for quantum information processing.

## 3. Outlook and conclusion

Given the fact that most quantum photonic applications strive for device miniaturization and compactness, where smaller devices usually deliver faster operation speeds and lower power consumption – here we discuss the potential impact where PMN-PT can have in the field of integrated quantum photonics. The combination of the high refractive index of PMN-0.38PT and the technological readiness of thin-film pulsed laser deposition[43] make such materials highly desirable for integrated photonic circuits constructed of dielectric/PMN-PT/dielectric stacks where photolithographically defined waveguides written into a PMN-PT slab core can achieve higher integration density, and ultimately high intensity optical field confinement for stronger nonlinear optical interaction. Furthermore, implantation of rare-earth ions, like $Er^{3+}$ into PMN-PT waveguides will allow for on-chip integration of NIR pump lasers for on chip PDC and circuit operation even at ambient temperatures[44]. The important electro-optical (EO) and electro-mechanical functionality of PMN-0.38PT makes it superior to widely used $LiNbO_3$[31], with the figure of merit half-wave voltage $V_\pi$ factor of 3 better, allowing lower operating voltages and therefore more efficient, compact and faster EO phase controlled on-chip Mach-Zehnder interferometers as well as Pockels cells[45,46]. It is important to mention that PMN-0.35PT has almost twice stronger $r_c$ of 81 pm/V which reduces the $V_\pi$ to just 452 V [31][47] in comparison to 2800 V typically found in $LiNbO_3$[48]. These characteristics make this material a very interesting object of investigation for smart applications in quantum photonics and beyond, where PDC photon sources, waveguide networks, and EO phase control and/or Pockels cells can be monolithically integrated on a single chip, with a capacity to extend the wavelength of mid-IR single photon sources for a range of novel quantum-enhanced photonic applications. We also note that since the dispersion parameters depend on the growth method and the doping levels, for experimental realization one would require to verify the exact dispersion parameters for a particular PMN-PT crystal.

In conclusion we have performed numerical investigation of the promising class of the functionally poled material PMN-PT and have identified GVM conditions for pure single photon generation at mid-IR spectral region. Our study brings a new class of functional materials into the quantum photonics research field. The suggested mechanism for active control of single photon spectral encoding possesses far reaching implications enabling accessing key characteristics responsible for single photon purity and indistinguishability, indispensable in many current and future quantum information processing and free-space quantum communication requiring reconfigurable operation of single photon sources.

**Acknowledgements**

The authors would like to acknowledge funding from Engineering and Physical Sciences Research Council (EPSRC) EP/N002962/1, RAM would like to acknowledge Heriot-Watt University fellowship, and FG acknowledges studentship funding from the EPSRC under EP/L015110/1 grant. The authors are also indebted to prof. Peter G.R. Smith for useful discussions.

**Table 1.** Comparison of some PMN-PT key properties against other nonlinear materials used for PDC. $T$ is the transparency range; $E_c$ is the coercive field; $T_c$ is the Currie temperature, $n$ is the refractive index at 633 nm; $r_c$ is the effective EO coefficient; $V_\pi$ is the reduced half-wave voltage calculated using $V_\pi = \lambda / r_c n^{-3}$ at $\lambda = 633$ nm[49]; $d^{2\omega}_{33}$ is the 2nd order nonlinear coefficient at 1064 nm.

| Material | T μm | $E_c$ kV/mm | $T_c$ °C | n | $V_\pi$, kV | $r_c$ pm/V | $d^{2\omega}_{33}$ pm/V |
|---|---|---|---|---|---|---|---|
| PMN-0.38PT | 0.45 – 6[26] | 0.4[27] | 180 | 2.6 | 0.875 | 41[31] | (a) |
| KTiOPO$_4$ | 0.4 – 4.5 | 2.5 | 950 | 1.76 | 4.05 | 28.6[50] | 16.9[51] |
| LiNbO$_3$ | 0.4 – 5 | 20 | 1145 | 2.29 | 2.6 | 20.1 | 27.5[52] |

a) Due to the absence of experimental values the 2nd order nonlinear coefficient for PMN-0.38PT can be estimated using method of F. Wang[52]; which gives a factor of 2 bound approximation. For the birefringence input parameters for PMN-0.38PT[36], KTiOPO$_4$[53,54] and LiNbO$_3$[55] it gives $d^{2\omega}_{33}$ of 12.6, 29 and 27.7 pm/V for these materials respectively.

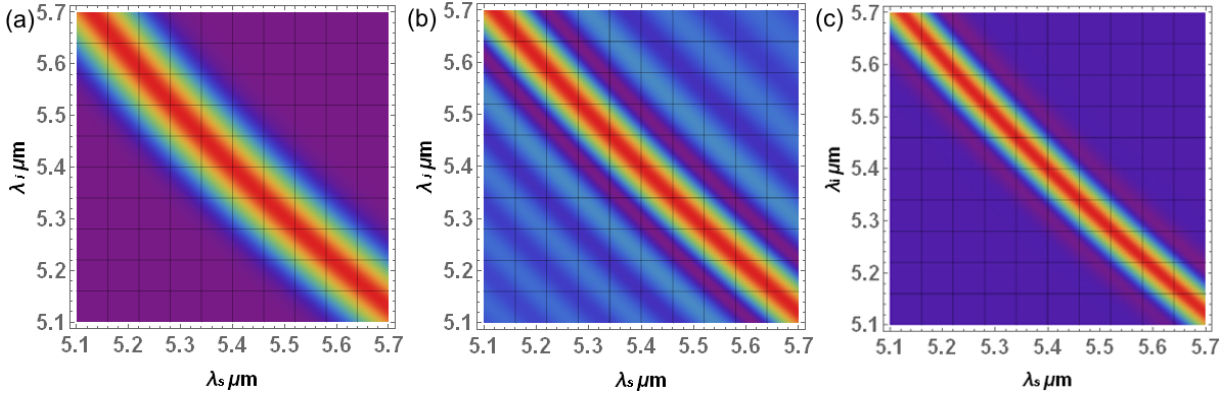

**Figure 1.** Type 0 ($e \rightarrow e+e$) density plots as a function of signal and idler photon wavelengths for (a) PEF inclined at -45° which reflects energy conservation; (b) PMF inclined almost identically to PEF thus giving no useful JSA for a defined bi-photon PDC state generation (c).

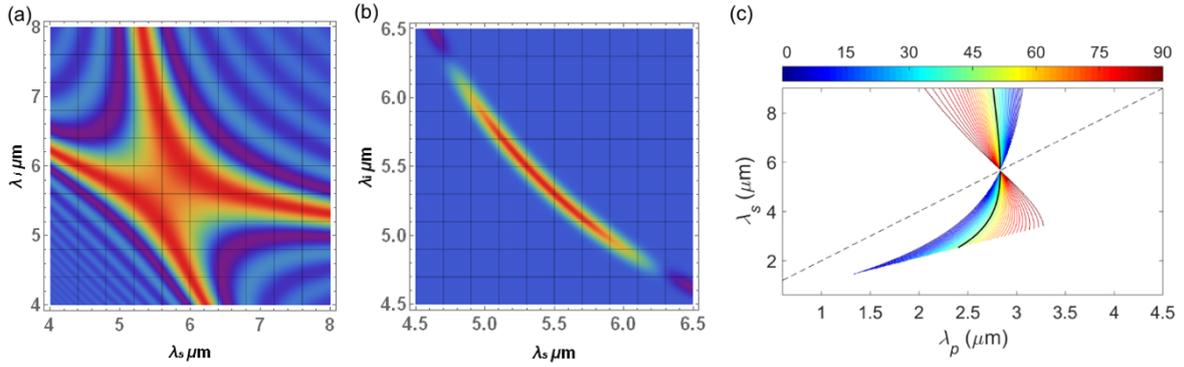

**Figure 2.** GVM results plots for type I ($e \rightarrow o+o$) (a) PMF showing singularity with a large crossover; (b) resulting JSA using PEF from Figure 1(a), and (c) spectral plots of the for optimal PMF inclination angle. The x- and y-axes in (c) represent pump and the corresponding signal photon wavelengths respectively; the diagonal dashed line indicates the degeneracy point at which signal and idler are spectrally symmetric. The black solid line indicates spectral configurations with i.e. where the highest signal-idler indistinguishability can be attained at θ = 45° i.e. when PMF aligned antidiagonal to the PEF.

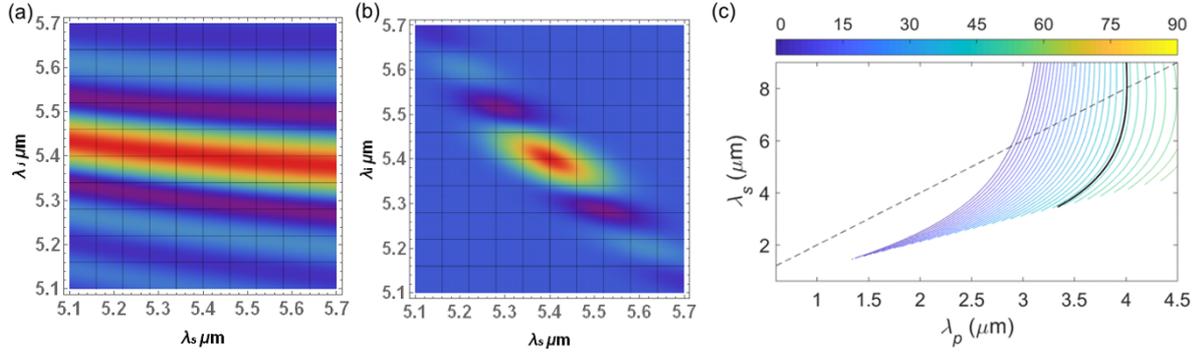

**Figure 3.** GVM results in signal/idler spectral space for type II ($e \rightarrow o+e$) PMF (a) exhibiting inclination angle of close to 0 degrees, which enables achieving high purity bi-photon JSA at 5.4 μm (b) with PEF from Figure 1(a), and (c) shows spectral positions for the optimal PMF inclination angles, also including non-degenerate cases.

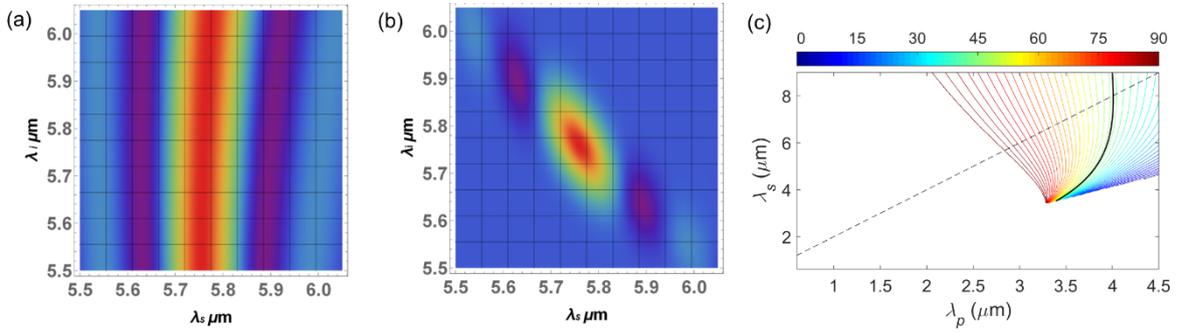

**Figure 4.** (a-b) GVM results in signal/idler spectral space for type II ($e \rightarrow e+o$); PMF (a) exhibiting inclination angle of exactly 90 degrees, which also enables achieving high purity bi-photon JSA at 5.76 μm (b) with PEF from Figure 1(a), and (c) shows spectral dependence for the all PMF inclination angles, also including non-degenerate cases.

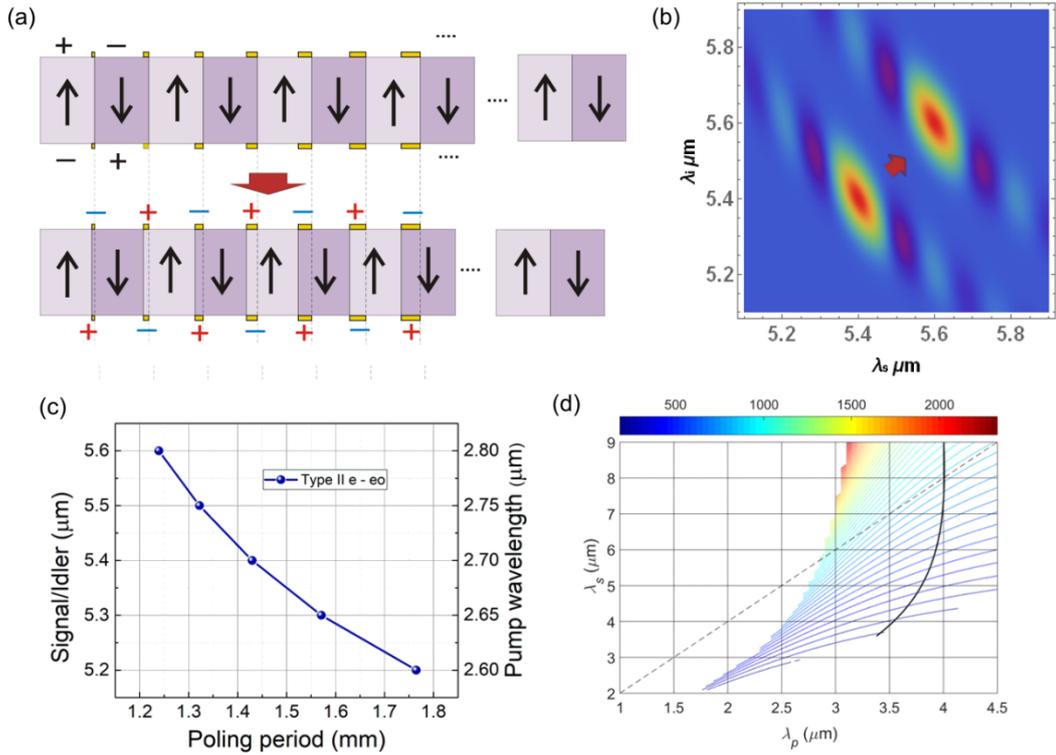

**Figure 5.** (a) Illustration of the poling period switching mechanism by the use of the applied electrode patterns to the pp-PMN-PT crystal. The crystal poled domains are depicted in different shades with direction of the spontaneous polarization indicated by the arrows. The electrode mask (yellow bars at each side of the crystal) is used for flipping the domain fraction in order to modify the poling period of the entire crystal. (b) corresponding JSA's for the two cases of the poling periods in (a). (c) shows the dynamic range of the center wavelength of signal/idler photons as a function of the modified poling period for the degenerated PDC; and (d) describes full range including for nondegenerate type II ($e \rightarrow e+o$) PDC cases, where the color lines represent the poling period $\Lambda$ (top bar in microns) for the entire space of pump/signal wavelength ranges.